\documentclass[twocolumn]{aa}
\usepackage{graphicx}
\usepackage{natbib}
\bibpunct{(}{)}{;}{a}{}{,}
\begin{document}
\title{Evidence of polarisation in the prompt gamma-ray emission from GRB\,930131 and GRB\,960924.}

\titlerunning{The BATSE Albedo Polarimetry System (BAPS)}

\author{D.R. Willis \inst{1,}\inst{2,}\inst{3}
\thanks{\emph{Present address:} INTEGRAL Science Data Centre (ISDC),
  Chemin D'Ecogia 16, CH-1290 Versoix, Switzerland. /
  Max-Planck-Institut fur extraterrestrische Physik, Postfach 1312
  D-85741 Garching bei Munchen, Deutschland.}
\and E.J. Barlow\inst{1}
\and A.J. Bird\inst{1}
\and D.J. Clark\inst{1}
\and A.J. Dean\inst{1}
\and M.L. McConnell\inst{4}
\and L. Moran\inst{1}
\and S.E. Shaw\inst{1,}\inst{3}
\and V. Sguera\inst{1}}
\institute{School of Physics and Astronomy, University of Southampton, SO17 1BJ, UK
\and Max-Planck-Institut fur extraterrestrische Physik, MPI, Garching,
Munich, Germany
\and INTEGRAL Science Data Centre, CH-1290 Versoix, Switzerland
\and Space Science Center, University of New Hampshire, Durham, NH 03824, USA}

\offprints{D.R. Willis, \email{David.Willis@obs.unige.ch}}

\date{Received now / Accepted then}

\abstract{The true nature of the progenitor to GRBs remains elusive; one characteristic that would constrain our understanding of the GRB mechanism considerably is gamma-ray polarimetry measurements of the initial burst flux. We present a method that interprets the prompt GRB flux as it Compton scatters off the Earth's atmosphere, based on detailed modelling of both the Earth's atmosphere and the orbiting detectors. The BATSE mission aboard the \textit{CGRO} monitored the whole sky in the 20\,keV -- 1\,MeV energy band continuously from April 1991 until June 2000. We present the BATSE Albedo Polarimetry System (BAPS), and show that GRB\,930131 and GRB\,960924 provide evidence of polarisation in their prompt flux that is consistent with degrees of polarisation of $\Pi>35$\% and $\Pi>50$\% respectively. While the evidence of polarisation is strong, the method is unable to strongly constrain the degree of polarisation beyond a systematics based estimation. Hence the implications on GRB theory are unclear, and further measurements essential.

}

\maketitle

%

\section{Introduction}
Whatever the mechanism, polarisation in the prompt gamma-ray flux from GRBs is evidence of strong magnetic fields within the burst. Theories on the GRB production mechanism can be constrained by different degrees of linear polarisation. For large degrees of polarisation, $\Pi \approx 80$\%, either shock accelerated synchrotron emission or a tuned Compton-drag model is the most likely \citep{lazzati, coburn, lyutikov}. For intermediate degrees of polarisation ($ 20$\%$< \Pi < 60 $\%) two electromagnetic models have emerged that involve either synchrotron emission as the dominant source of radiation or as the result of viewing the burst from just outside the edge of the jet \citep{granot,ghisellini}. Low degrees of polarisation can be a result of either a hydrodynamic model or from flux with a higher degree of polarisation experiencing partial depolarisation. 

The high degree of linear polarisation initially reported in the prompt flux of GRB\,021206 [degree of polarisation, $\Pi = 80 \pm 20$\%, \citet{coburn}] with the \textit{RHESSI} experiment \citep{mcconnell2} has stimulated much interest in the implications this has on GRB theory. Though GRB astronomy has the advantage of large fluxes, any polarimetric measurements are still dominated by systematic effects that can only be properly quantified by careful modelling. The importance of correctly evaluating the systematic effects is paramount in any measurement of GRB polarisation as was emphasised by \citet{wigger} in the thorough re-analysis of the initial \textit{RHESSI} result \citep{coburn}. \citet{wigger} reanalysed the degree of polarisation to be $\Pi = 41^{+57}_{-44}$\%. This implies that many of the production mechanism theories are now similarily competitive. 

Besides the \textit{RHESSI} instrument \citep{lin}, there are several other instruments that could be suitable gamma-ray polarimeters: IBIS and SPI aboard \textit{INTEGRAL} \citep{winkler} have the as yet untested capability of detecting gamma-ray polarisation; COMPTEL \citep{schoenfelder} and BATSE, both aboard the \textit{Compton Gamma-Ray Observatory} (\textit{CGRO}) \citep{fish89a} are also possible candidate instruments. For Compton polarimetry, the instrument is required to scatter the photons and detect the modulation induced by the polarisation-sensitive differential cross-section for Compton scattering (equation \ref{eqn:kn}), see \citep{lei2} for a review of Compton polarimetry in gamma-ray astronomy.

\begin{equation}
\frac{d\sigma}{d\Omega} = \frac{r_{0}^{2}}{2} \left( \frac{E'}{E_{0}}
 \right)^{2}\left( \frac{E'}{E_{0}} + \frac{E_{0}}{E'} - 2sin^{2}\theta cos^{2} \phi \right)  
\label{eqn:kn}
\end{equation}

where $r_{0}^{2}$ is the classical radius of the electron, $E_{0}$ the energy of the incident photon, $E'$ the energy of the scattered photon, $\theta$ the scatter angle and $\phi$ the azimuthal angle relative to the polarisation direction. A problem in both COMPTEL and \textit{RHESSI} is that each instrument has to both scatter and detect the linearly polarised photons. The scattering \emph{effective area}, $A_{eff}$, is small, resulting in a reduction in the number of double events required to detect the polarisation-induced azimuthal modulation. COMPTEL also had the problem that to act as a polarimeter the double-event rate had to be quite high \citep{lei1}, and at these high event rates vital kinematic information was lost in the telemetry. This is also expected to affect the polarimetry mode in IBIS to some extent.

To both scatter and detect the polarisation-induced azimuthal modulation, an instrument has to be either highly event selective and tracking sensitive or requires a large scattered flux, larger than the majority of GRBs can supply to \textit{RHESSI} and \textit{INTEGRAL} \citep{swinyard}. The problem of tracking and selecting the events most efficiently is currently being addressed in missions such as the \textit{Tracking and Imaging Gamma-Ray Experiment} (\textit{TIGRE}) \citep{oneill} and the \textit{Gamma-RAy Polarimeter Experiment} (\textit{GRAPE}) \citep{mcconnell3}. BATSE however is in the unique position to address the problem of increasing $A_{eff}$ for scattering the incident gamma-ray photons. \citet{mcconnell1} proposed that polarimetry was possible by combining the large $A_{eff}$ of the Earth's atmosphere for Compton scattering with the $4\pi$ steradian field-of-view of the eight BATSE modules. Unfortunately the response of the \textit{CGRO} to non-planar flux was difficult to handle analytically due to the uncollimated nature of BATSE and so could only be properly evaluated with modelling techniques. The modelling technique, called Mass Modelling, has been developed at Southampton and has been employed successfully on BATSE (\citet{mattsthesis,shaw1,shaw4} and references therein), \textit{INTEGRAL} \citep{ferguson} and \textit{Swift} \citep{davesthesis}. The complex analytical situation and systematic effects resulting from immersing the \textit{CGRO} in the GRB flux scattered off the Earth's atmosphere can be modelled using this Mass Modelling approach, enabling polarimetry measurements with BATSE to take full advantage of the large $A_{eff}$ that the Earth's atmosphere provides.

Here we present the technique of albedo Compton polarimetry from the Earth's Atmosphere \citep{mcconnell1} combined with the BATSE Mass Model (BAMM) \citep{shaw1}, the BATSE Albedo Polarimetry System (BAPS).

\section{The BATSE Albedo Polarimetry System (BAPS)}
\label{sec:baps}
\begin{figure*}[ht]
\centering
\includegraphics[width=17cm]{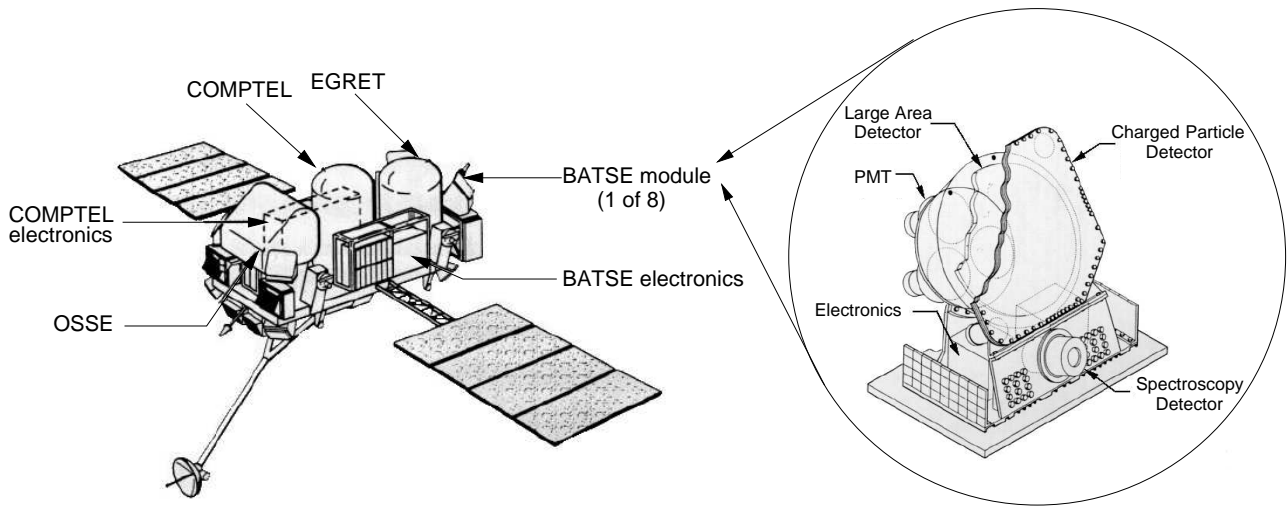}
\caption{The Compton Gamma-Ray Observatory (\textit{CGRO}) and one of the 8 BATSE detector modules (drawings courtesy of the National Space Science and Technology Centre, Huntsville, AL, USA).}
\label{fig:cgro}
\end{figure*}

The BATSE mission aboard \textit{CGRO} monitored the whole sky in the 20\,keV -- 1\,MeV energy band continuously from April 1991 until June 2000. It was designed primarily to detect and locate GRBs \citep{fish89b}. BATSE consisted of eight uncollimated 2025cm$^{2}$ NaI(Tl) Large Area Detectors (LADs), which had a combined field of view of $4\pi$ steradians. Also positioned below each LAD was a Spectral Detector (SD) made of 127 cm$^{2}$ NaI(Tl). The SDs were optimized for energy coverage and resolution, and were operated over the range 10\,keV -- 100\,MeV.  An illustration of the \textit{CGRO} and a BATSE module can be seen in Figure \ref{fig:cgro}. BATSE was an extremely successful experiment and the current burst catalogue has over 8000 triggers ({\small http://cossc.gsfc.nasa.gov/batse/BATSE\_Ctlg/index.html}). The discovery that GRBs are located at cosmological distances is largely due to data collected with BATSE \citep{meegan}. 

\begin{figure*}[ht]
\centering
\includegraphics[width=17cm]{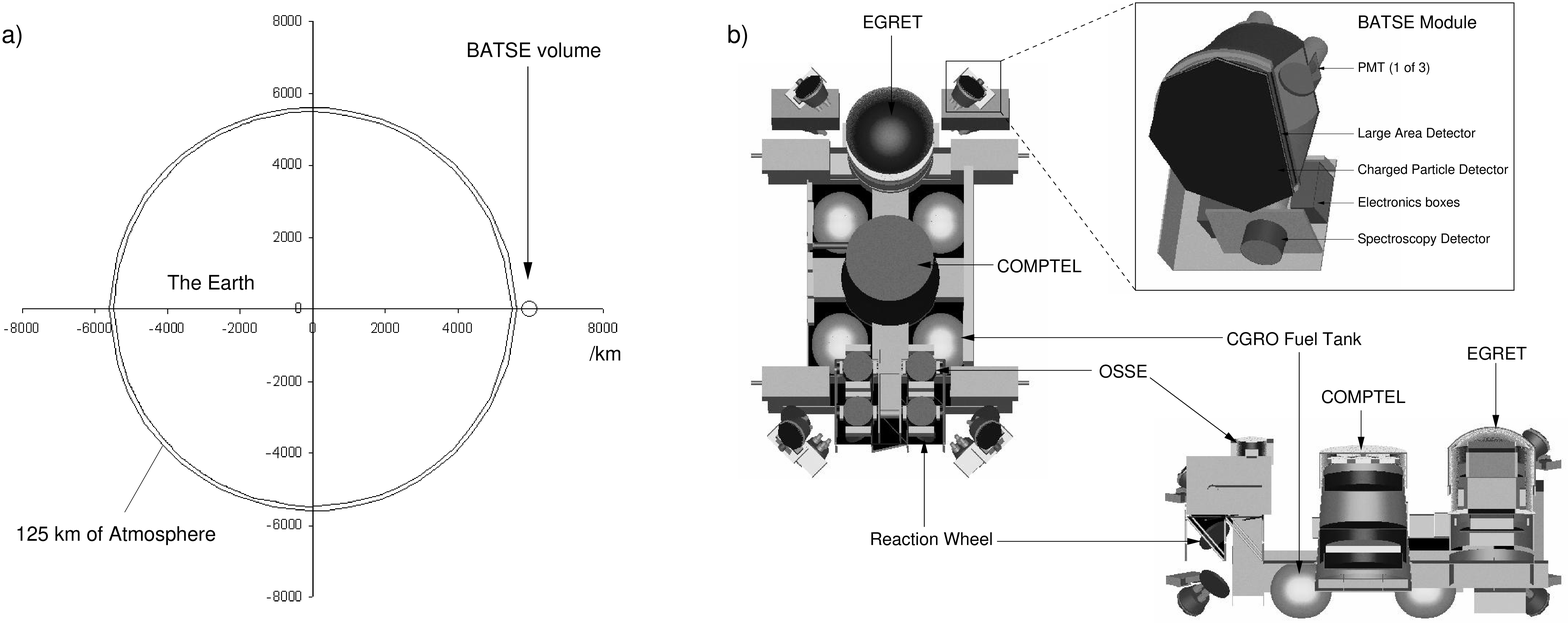}
\caption{{\bf (a)} This shows the relative scale of the Earth, its atmosphere (up to 125 km) and the virtual BATSE volume for the GEANT4 polarisation sensitive simulations.{\bf (b)} BAMM contains detailed structural, geometrical and material infomation for BATSE, EGRET, COMPTEL, OSSE and the \textit{CGRO} spacecraft. Cut-aways of the BATSE module (top right) and the whole \textit{CGRO} (bottom right) show the internal detail. To illustrate the crucial components that will deflect or absorb gamma-ray photons, some external volumes have been made transparent.}
\label{fig:albedo}
\end{figure*}

BAPS uses both the BATSE Mass Model (BAMM) and a model of the Earth's atmosphere. The system is highly CPU intensive and combining both models into one model is impractical due to the relative scales. The model of the Earth's atmosphere was written in GEANT4 \citep{geant}, in order to include the polarised Compton scattering process. It consists of a ball of radius 6503 km [Earth radius + 125 km] made of an integrated atmospheric composition of constant density [$\rho = 8.33 \times 10^{-5}$ g.cm$^{-3}$, 38\% O$_{2}$ : 62\% N$_{2}$], the effects of this assumption are discussed in section \ref{sec:atmos}. The BATSE volume was an imaginary sphere of radius 200 km at an altitude of 500 km. Figure \ref{fig:albedo} shows the scale of the Earth albedo model alongside a schematic of BAMM. BAMM \citep{shaw1} was written in GEANT3 to dynamically flat-field the variable background for the purposes of occultation imaging \citep{harmon} and the subsequent production of an all-sky survey \citep{shaw4}. BAMM reproduces the instrument performance to a level of accuracy that is impossible through analytical techniques alone. For a full review of the mass modelling approach and its applications, see \citet {dean1}.

BAPS simulates the GRB flux as it scatters off the atmosphere; the
distribution of this scattered flux is recorded as it passes through a
volume equivalent to where the \textit{CGRO} was at the time of the
burst. Polarised flux preferentially scatters perpendicular to the
direction of the polarisation vector, so any distribution produced as
a result of polarised flux will appear as an anti-phased excess
towards the limbs of the Earth, when viewed from the direction of the
burst. The recorded distribution scattered from the atmosphere is then
used as an input to the BAMM Monte Carlo and the count-rates simulated
for each LAD. This process is carried out for an unpolarised burst and
for a fully polarised burst, for a range of polarisation angles. The
simulated LAD count-rates are then compared to the burst count-rates
in the BATSE data. The fitting procedure varies the percentage
polarisation and the polarisation angle. The Z statistic, defined by
$Z=\frac{R_{1}-R_{2}}{\left(\frac{R_{1}}{t_{1}}+\frac{R_{2}}{t_{2}}\right)^{\frac{1}{2}}}$
where R and t are the count-rates and observation times respectively, is used to
compare the two count-rates: simulated and recorded. 

BATSE uses two formats of data, CONTinuous (CONT) and DISCriminator Large Area (DISCLA). DISCLA data contains 4 broad energy channels with a binning width of 1.024\,s. CONT data contains 16 energy channels with a binning width of 2.048\,s. Due to BAPS only being sensitive to the Compton domain and below, CONT data is used to optimise this sensitivity. The sacrifice of better timing resolution is not a problem until time-evolving polarimetry is considered.

\section{Burst Selection: GRB\,930131 and GRB\,960924}
\label{sec:selection}

\begin{figure}[ht]
\centering
\includegraphics[width=8cm]{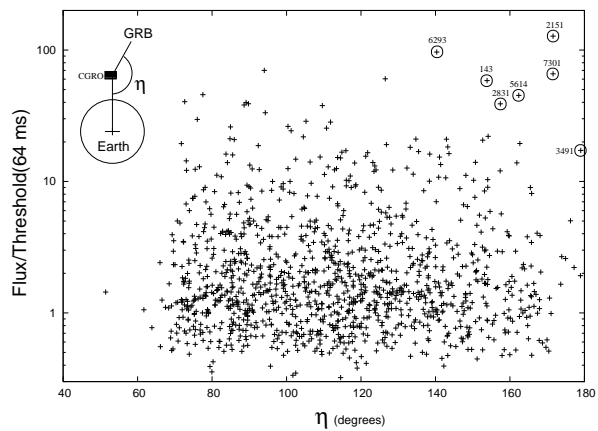}
\caption{All BATSE triggers plotted to identify the six most suitable candidates for polarimetry. Triggers 2151 and 5614 were initially selected with the remainder discarded due to their particularly long and structured lightcurves.}
\label{fig:64}
\end{figure}

The selection criteria for a suitable burst are that the burst has to be short, strong, have a geocentre to burst angle, $\eta$, as close to 180$^{o}$ as possible (i.e. to have BATSE directly between the Earth and the burst) and for it to have no obvious evolving features. This selection reduces any background variations, maximise the polarisation-induced modulations and avoid the possibility of reprocessing at the source reducing the degree of polarisation, perhaps due to reverse shocks and scattering at the source. Suitable bursts were identified via a plot of burst strength against burst to geocentre angle, $\eta$ (Figure \ref{fig:64}). Two of the strongest and most suitably placed BATSE triggers were identified as being suitable for this study: trigger 2151 (GRB\,930131) and trigger 5614 (GRB\,960924). The remaining triggers were either too long or had clear evolving features in the light curve, and so were discarded at this stage. The parameters of these two selected bursts are displayed in table \ref{GRBs}. 

\begin{table}
\begin{center}
\caption{A table of the relevant parameters to BAPS. The spectral parameters used were obtained using a standard spectral characterisation \citep{band}.}
\label{GRBs}
\begin{tabular*}{20pc}{lll} \hline\hline 
GRB & {\it 930131} & {\it 960924}  \\ \hline
Trigger & 2151 & 5614 \\
$\eta^{\circ}$ & 170.7 & 162.4 \\
CONT bins (2.048 seconds) & 2 & 4 \\
Band model $\alpha$ & -1.2 & -0.3 \\
Band model $\beta$ & -2.5 & -9.9 \\
Band model $E_{break}$ (keV) & 439 & 251 \\ \hline
\end{tabular*}
\end{center}
\end{table}

\section{Analysis}
\label{sec:analysis}

\subsection{Simulations}
\begin{figure}[ht]
\centering
\includegraphics[width=9cm]{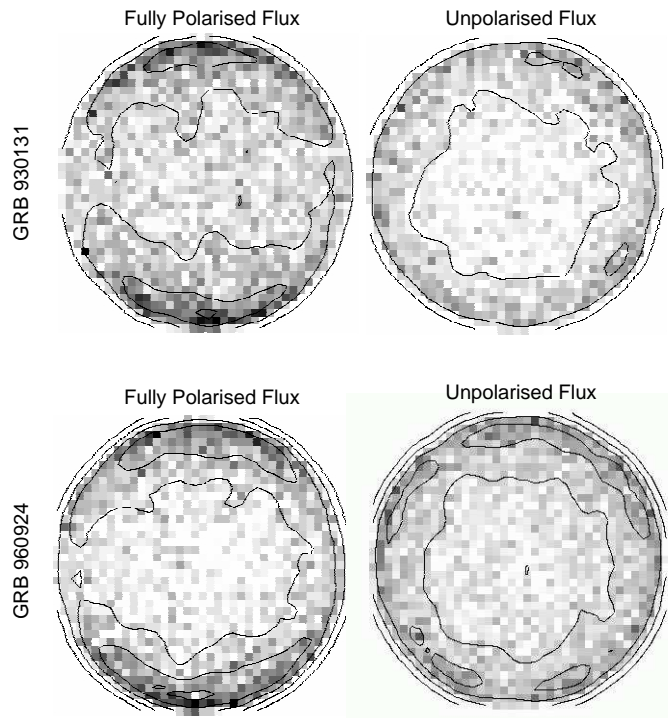}
\caption{The simulated angular distribution of gamma-ray flux
  reflected off the Earth's atmosphere during GRB\,930131 and
  GRB\,960924 for 100\% polarised and unpolarised fluxes. Contours are
  the 0, 5, 10, 15, 20 and 25 mean counts per pixel
  (4.5$^{\circ}\times$4.5$^{\circ}$) for 5$\times$10$^{7}$ incident
  GRB photons. This pixelated distribution is used as the input to BAMM.}
\label{fig:albdist}
\end{figure}

The first stage of BAPS is to produce the input distributions for the Mass Model. Fully polarised flux from the GRBs (with polarisation angle incremented every $15^{o}$), and an unpolarised burst flux were simulated incident on the atmosphere. 0.03\% of the incident flux was seen to reflect and pass through the imaginary bubble containing the \textit{CGRO}. The albedo angular distribution for polarised and unpolarised flux for both the GRBs can be seen in Figure \ref{fig:albdist}.

\begin{figure}[ht]
\centering
\includegraphics[width=7.5cm]{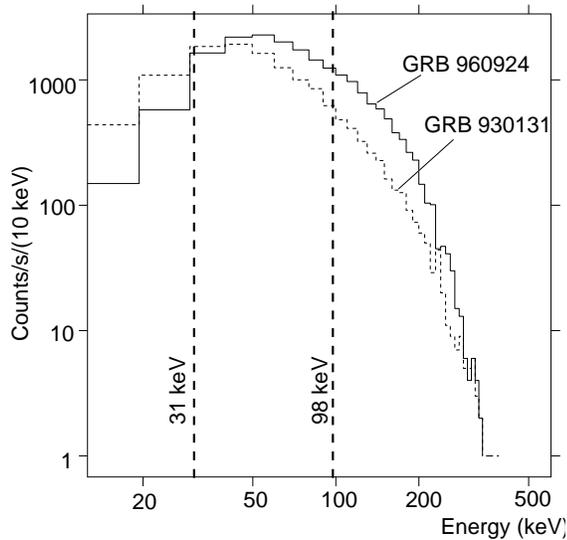}
\caption{The spectra of the albedo component of GRB\,930131 and GRB\,960924. The dotted lines represent the domain in which the data is compared: 31--98\,keV (LAD CONT channels 3--6). These limits equate to 67\% and 62\% of the albedo flux respectively.}
\label{fig:albspec}
\end{figure}

BAMM was then used to simulate the effect these distributions of albedo flux have on the count-rates in the LADs during each burst. The direct GRB flux is also a simulated component. As seen in the albedo spectra (Figure \ref{fig:albspec}) this technique is only sensitive in the Compton domain and so the energy range of the fitting is restricted to LAD CONT channels 3--6 (31--98\,keV). The Compton domain extends up to higher energies but these restrictions equated to 67\% and 62\% of the albedo flux for GRB\,930131 and GRB\,960924 respectively.

\subsection{LAD selection and fitting}
The LADs selected are divided into two types, the first are used to normalise the model count-rates for the direct and albedo fluxes and the second are LADs pointed such that the albedo count-rate varies considerably depending on the degree and angle of polarisation. The remaining LADs are ignored if they have low simulated count-rates and low levels of polarisation-induced modulation, thus reducing the systematic and counting errors in the system and avoiding adding unconstrainable parameters into the fitting process. The component direct from the burst is normalised using the count-rate from the most burstward LAD. The simulated count-rates are fitted to the recorded count-rates with the normalisation of the direct flux fixed and the albedo flux, degree and angle of polarisation as free parameters. The ambient background is removed by taking a mean count-rate from before and after the burst. The fit statistic used is a simple Z-test for comparing the simulated and recorded count-rates.

The criteria for selecting the LADs is best demonstrated by looking at the simulated count-rate components for each LAD, see tables \ref{GRB93} and \ref{GRB96}. The figure-of-merit consists of the percentage of simulated albedo flux, {\bf A}, in each LAD compared to the total flux, the maximum variation due to GRB polarisation angle and the level of albedo flux compared to the LAD with the maximum albedo flux. For each burst one LAD is selected to normalise the direct burst flux, one to assist in the normalisation of the albedo flux and two to measure polarisation-induced modulation. For different bursts the number of LADs selected may vary according to their figure-of-merit.

\subsubsection{GRB\,930131}
The LADs selected are \textit{LAD1}, \textit{LAD2}, \textit{LAD3}, and \textit{LAD4}. \textit{LAD1} and \textit{LAD2} have the largest albedo count-rate and are most sensitive to polarisation. \textit{LAD4} is used to set the normalisation for the direct flux and \textit{LAD3} is used to assist the normalisation of the albedo flux. 

\begin{table}
\begin{center}
\caption{GRB\,930131 LAD figure-of-merit. The LADs used in the fitting are marked in bold.}
\label{GRB93}
\begin{tabular*}{20pc}{cccc} \hline\hline
LAD & $\frac{A}{A+Direct}$ \% & Var $\left(\frac{A_{pol}-A_{unpol}}{A_{unpol}}\right)$ \%  & $\frac{A}{A_{max}}$ \% \\ \hline
LAD0 & 1   & 45 & 38  \\ 
{\bf LAD1} & 49  & 38 & 59  \\ 
{\bf LAD2} & 70  & 17 & 60  \\
{\bf LAD3} & 73  & 7  & 100 \\
{\bf LAD4} & $<$1 & 58 & 1 \\
LAD5 & 16  & 25 & 2 \\ 
LAD6 & $<$1 & 38 & 9 \\
LAD7 & 9   & 31 & 17 \\ \hline
\end{tabular*}
\end{center}
\end{table}

\subsubsection{GRB\,960924}
The LADs selected are \textit{LAD0}, \textit{LAD2}, \textit{LAD4}, and \textit{LAD6}. \textit{LAD6} is used to set the normalisation for the direct flux. For this orientation of the \textit{CGRO}, the polarisation-induced variation appears in all LADs. \textit{LAD0} is used to primarily assist the normalisation of the albedo flux whereas \textit{LAD2} and \textit{LAD4} show the largest albedo flux and have $>$ 30\% variability due to polarisation angle and so are used to test any polarisation-induced modulation.

\begin{table}
\begin{center}
\caption{GRB\,960924 LAD figure-of-merit. The LADs used in the fitting proceedure are marked in bold.}
\label{GRB96}
\begin{tabular*}{20pc}{cccc} \hline\hline 
LAD & $\frac{A}{A+Direct}$ \% & Var $\left(\frac{A_{pol}-A_{unpol}}{A_{unpol}}\right)$ \%  & $\frac{A}{A_{max}}$ \% \\ \hline
{\bf LAD0} &  31   & 11 & 100  \\ 
LAD1 &      51   & 10 & 29  \\ 
{\bf LAD2} & 8    & 30 & 57  \\
LAD3 &      7   &  35  & 14 \\
{\bf LAD4} & 3    & 36 & 62 \\
LAD5       & 4   & 44 & 13 \\ 
{\bf LAD6} & $<$1 & 21 & 8 \\
LAD7       & 4    & 41  & $<$1 \\ \hline
\end{tabular*}
\end{center}
\end{table}

\subsection{Results}
\begin{figure*}[ht]
\centering
\includegraphics[width=15cm]{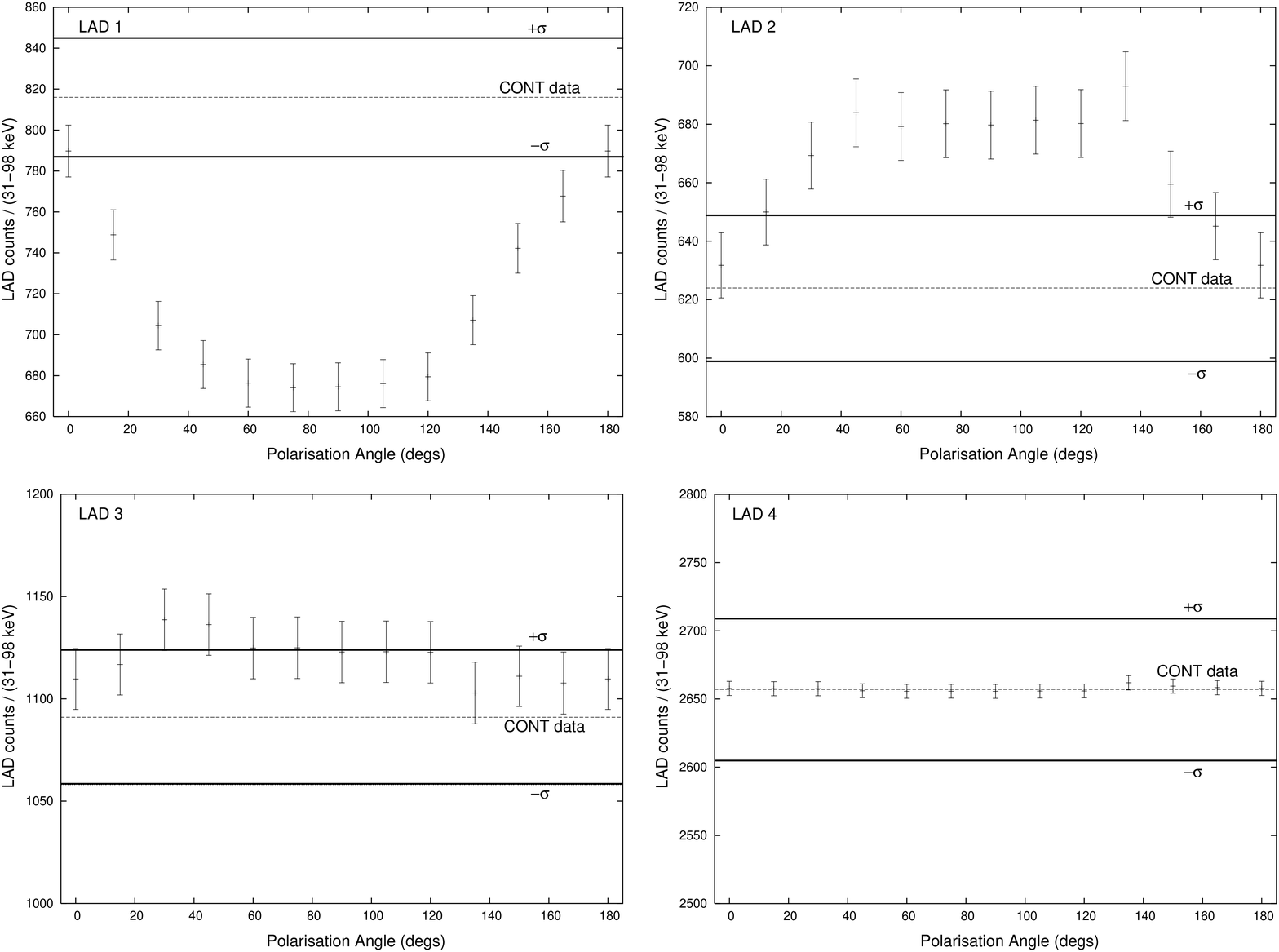}
\caption{A comparison of the fitted simulated count-rates with the LAD CONT data for GRB\,930131. The CONT data is shown as a dashed line, the error on that count is shown as solid lines. The best fit for each polarisation angle is shown with the relevant counting error from the Monte Carlo. \textit{LAD3} and \textit{LAD4} simulations are expected to match the recorded data, independent of polarisation. \textit{LAD1} and \textit{LAD2} are sensitive to polarisation in the GRB's prompt flux and will show an improvement in fit for angles that corresponds to the burst polarisation angle.}
\label{fig:4lad2151}
\end{figure*}

\begin{figure}[ht]
\centering
\includegraphics[width=8cm]{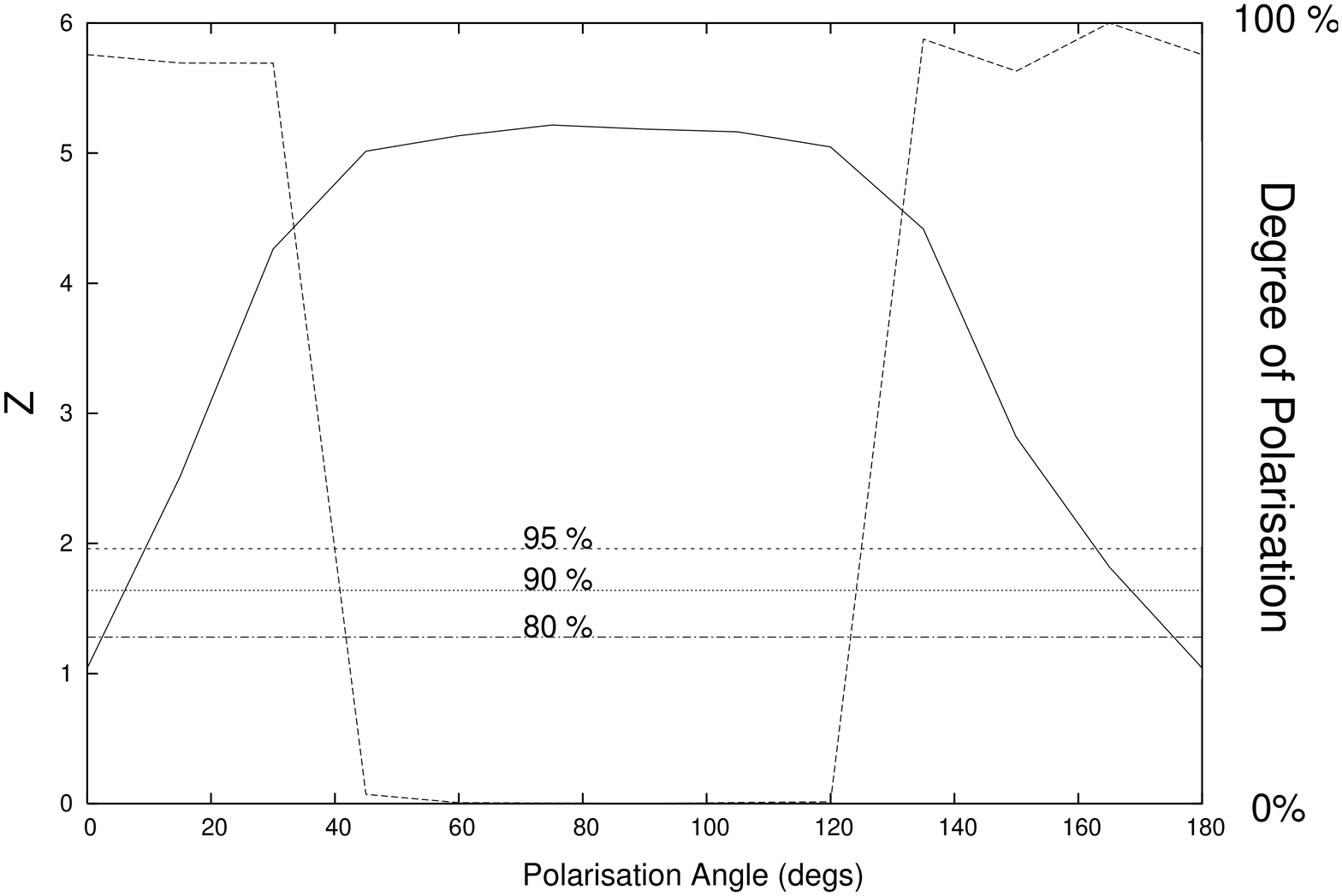}
\caption{For GRB\,930131, the Z statistic (solid line) for each of the best fits from Figure \ref{fig:4lad2151}. The statistic becomes significant (95\% confidence) between polarisation angles of 163$^{\circ}$ and 9$^{\circ}$ and this corresponds to a high ($>$90\%) degree of polarisation (dashed line). The Earth's North/South equivalent angle is at 5$^{\circ}$, see section \ref{sec:NS}.}
\label{fig:2151zdeg}
\end{figure}

\begin{figure*}[ht]
\centering
\includegraphics[width=15cm]{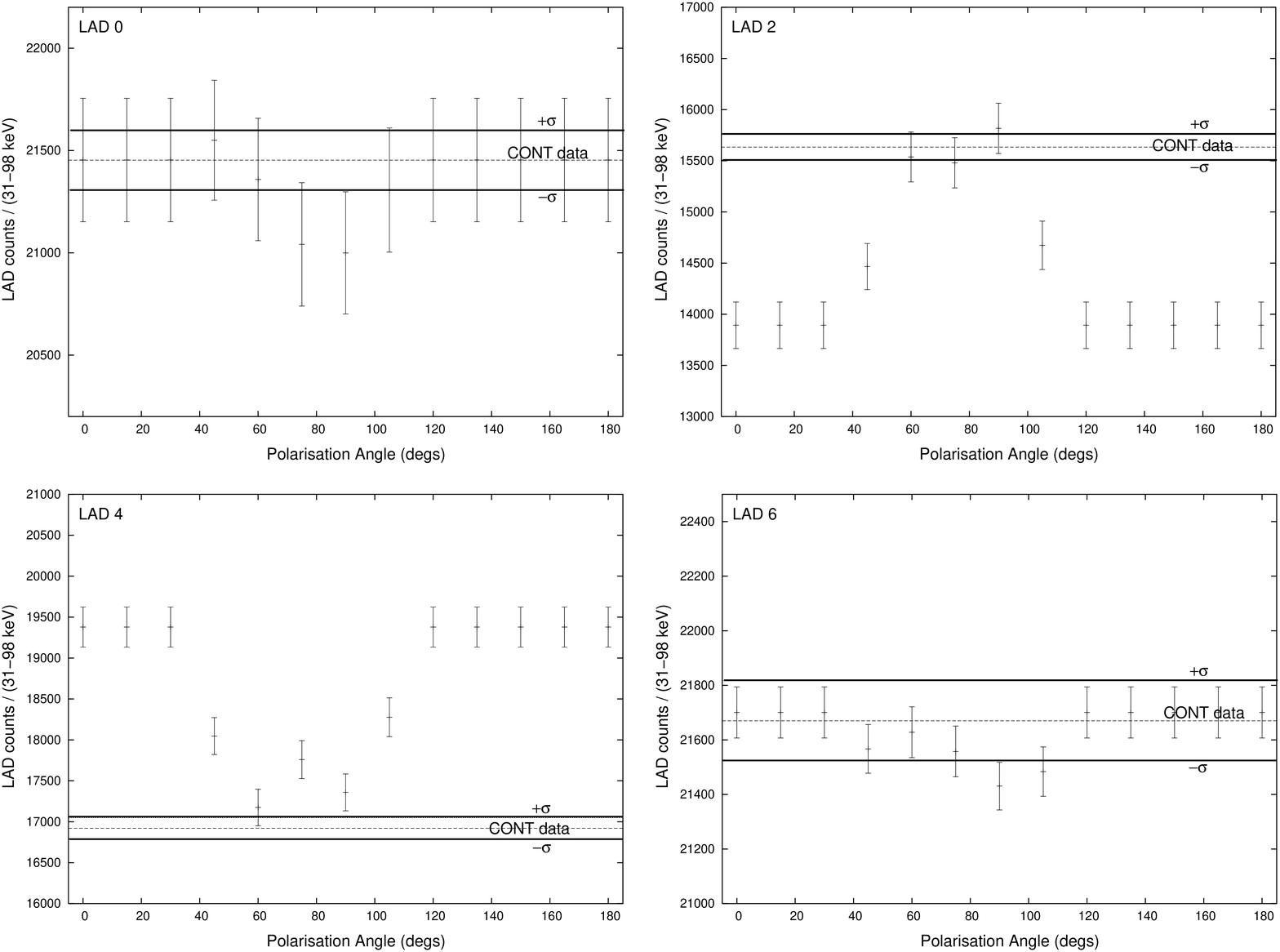}
\caption{A comparison of the fitted simulated count-rates with the LAD CONT data for GRB\,960924. The CONT data is shown as a dashed line, the error on that count is shown as solid lines. The best fit for each polarisation angle is shown with the relevant counting error from the Monte Carlo. \textit{LAD6} simulations are expected to match the recorded data, independent of polarisation. \textit{LAD2} and \textit{LAD4} are sensitive to polarisation in the GRB's prompt flux and will show an improvement in fit for the specific angle that the burst is polarised to. \textit{LAD0} is largely insensitive to polarisation and so should be consistent with the CONT data regardless of polarisation.}
\label{fig:4lad5614}
\end{figure*}

\begin{figure}[ht]
\centering
\includegraphics[width=8cm]{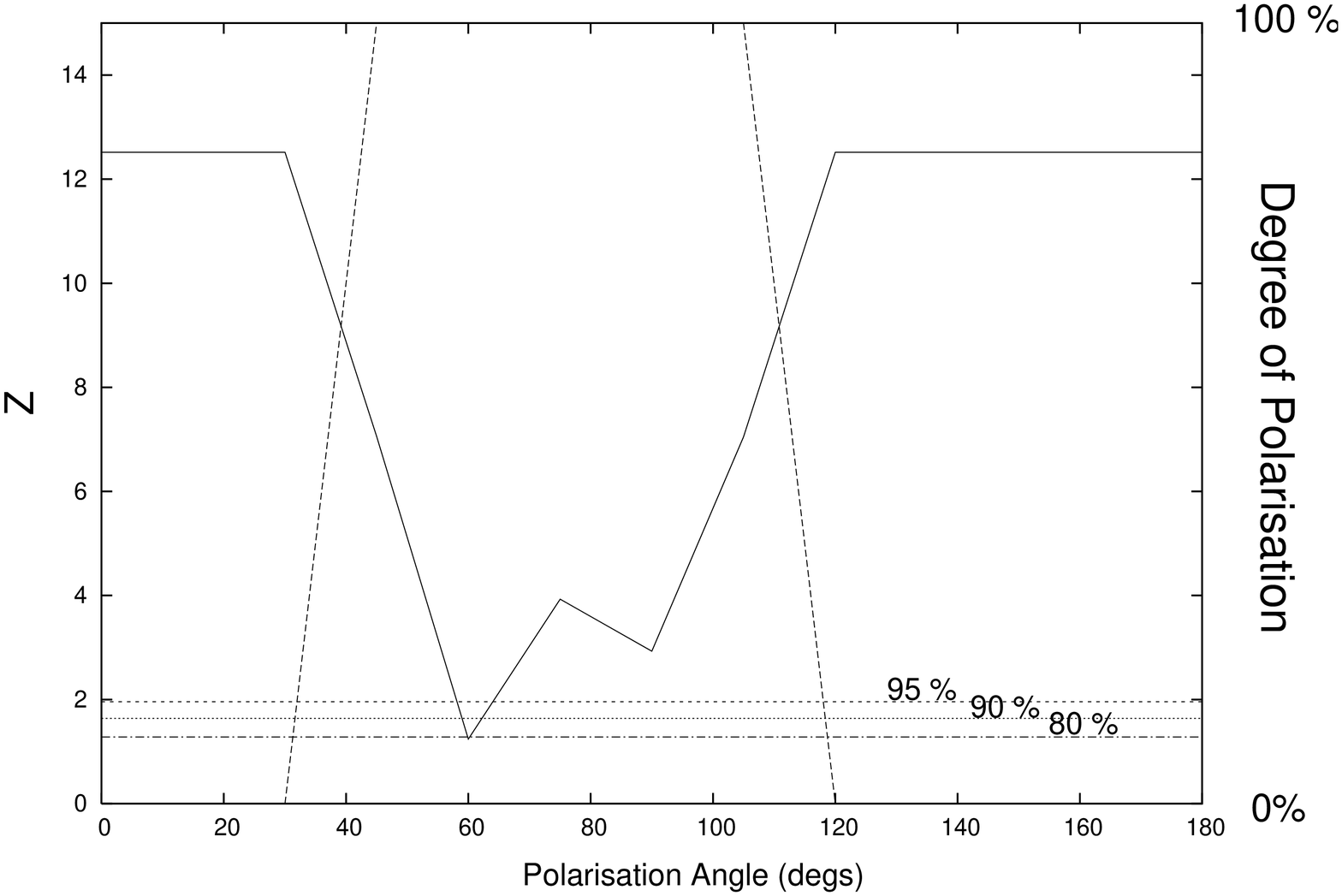}
\caption{For GRB\,960924, the Z statistic (solid line) for each of the best fits from Figure \ref{fig:4lad5614}. The statistic becomes significant (95\% confidence) between polarisation angles of 58$^{\circ}$ and 64$^{\circ}$ and this corresponds to a high (100\%) degree of polarisation (dashed line). The Earth's North/South equivalent angle is at 47$^{\circ}$, see section \ref{sec:NS}.}
\label{fig:5614zdeg}
\end{figure}

The selected LADs are simultaneously fitted to the CONT count-rates and the best fit for each incident polarisation angle is shown in Figures \ref{fig:4lad2151} and \ref{fig:4lad5614}. The significance of each of these polarisation angle fits can be taken from Figures \ref{fig:2151zdeg} and \ref{fig:5614zdeg}, the values for which were taken from a Z-test comparing the total count-rates in the LADs from the simulation and CONT data using the parameters supplied by the best fit for each polarisation angle. Also included in these plots of the Z statistic are the corresponding $\Pi$ values.

GRB\,930131 shows evidence of a high degree of polarisation. In this case, using simulated data with the polarisation angle at 90 degrees to the actual polarisation angle, the best fit will be to the unpolarised simulated flux, due to the phased nature of the excess at the Earth's limb. Hence for this burst the degree of polarisation will only move towards a higher value when the fit is improved by the polarised GRB. Though this is seen in Figure \ref{fig:2151zdeg} the sharp change in the degree of polarisation compared to the smoother change in the Z statistic suggests that this technique is not particularily sensitive to the degree of polarisation and can initially only be used as evidence of polarisation in the prompt GRB flux and that GRB\,930131 is best fit by a fully polarised model (though see section \ref{sec:estimate} for an estimate of $\Pi$).

GRB\,960924 also shows evidence of a high degree of polarisation. The variation in the Z statistic in this case shows a clear dip between 35$^{\circ}$ and 120$^{\circ}$ and is flat for the remainder of angles. The sharp increase in the degree of polarisation compared to the dip in the Z statistic clearly shows the angles for which a polarised burst fits the CONT data better than an unpolarised burst. Again this is evidence of polarisation, that is best fit by a fully polarised model.

\subsection{Systematic Effects}
\label{sec:sys}
These results are expected to contain a series of systematic effects and a discussion and evalution of these effects follows. The LAD selection by way of a figure-of-merit for each burst is expected to minimise these.

\subsubsection{North-South Albedo Gamma-Ray Excess}
\label{sec:NS}
The atmosphere itself is a strong source of gamma radiation. Cosmic
ray interactions in the atmosphere radiate gamma-rays
isotropically. This emissivity depends solely on the intensity of the
incident cosmic rays and the atmospheric depth \citep{dean2}. An
angle-averaged spectrum of this albedo flux is also presented by
\citet{gehrels91}. This albedo flux will be concentrated at the
Earth's limb as the largest portion of the atmosphere is being
observed. Along the limb, the strongest source of albedo flux will be
towards the polar regions, as the rigidity of the geomagnetic field
deflecting the cosmic rays is at its weakest \citep{smart}. This will
result in an excess that will be reminescent of the excess that BAPS
will be looking for.  During these bursts BATSE was at an inclination
of -16$^{\circ}$ and -10$^{\circ}$ for GRB\,930131 and GRB\,960924
respectively. The emission from the south pole would have been visible
to BATSE in both cases. The cosmic ray induced albedo flux would therefore falsely suggest a polarisation angle of 5$^{\circ}$ and 47$^{\circ}$ respectively, which corresponds to the inferred polarisation angle for GRB\,930131 but not for GRB\,960924. However, GRB\,930131 should not be discounted as a possible source of prompt polarised flux as any excess produced from the South pole should be removed during the background subtraction within the fitting proceedure. 

\subsubsection{East-West Cosmic Ray Anisotropy}
Another effect of the geomagnetic field is the cosmic ray east-west anisotropy, from the direct detection of cosmic ray particles. The rigidity of the geomagnetic field indicates the ability of the field to fully deflect a charged particle. Faraday's Law defines the direction that a cosmic ray is deflected in, and results in an excess in the westerly direction. This would correspond, in this case, to a falsely inferred polarisation angle of 95$^{\circ}$ and 137$^{\circ}$ for GRB\,930131 and GRB\,960924 respectively. These polarisation angles do not show an improvement in the fit and so this anisotropy is not responsible for a positive detection. Similar to the polar albedo excess discussed in section \ref{sec:NS}, any anisotropy should not be seen due to the background removal and it is not expected to change on the order of seconds. Moreover, any level of background due to cosmic ray interactions will be low as each BATSE LAD had a plastic scintillator vetoing cosmic ray events. Finally, it is estimated that both the North-South albedo excess and the East-West cosmic ray anisotropy are expected to be extended by 20--30$^{\circ}$, leaving few of the potential polarisation angles free from these systematic effects. For weaker bursts, these North-South and East-West regions may give false detections.

\subsubsection{Modelling Software}
\label{sec:atmos}
Combining two GEANT models will introduce a level of systematic error but due to the relative scale of the two models this is unavoidable. The GEANT 3 model of the \textit{CGRO} is proven to work to an acceptable level of accuracy \citep{shaw1}. However, the GEANT 4.6.1 model of the Earth's atmosphere may introduce some problems. GEANT 4 is a new suite of software and is still under development, requiring verification at each stage. One area that requires modification was highlighted by \citet{mizuno}: the modulation produced from scattering a polarised beam. For their GEANT model the unmodified version of GEANT4 was seen to scatter a Crab (25--200\,keV) spectrum with around 30\% variation with scattering angle, whereas the beam-test verified and modified version of GEANT was seen to give a modulation of around 64\%. The models are entirely different but this suggests that the level of modulation in each LAD due to polarisation in BAPS could be up to half of what it should physically be. This is undoubtably a systematic error which will degrade the sensitivity of BAPS but not one that would give a false indication of polarisation. Future runs will include the modifications suggested by \citet{mizuno} and it is expected that BAPS will become far more sensitive as a result, and will give a better indication as to the degree of polarisation.

\subsubsection{Constant Density Atmospheric Model}
\begin{figure}[ht]
\centering
\includegraphics[width=9cm]{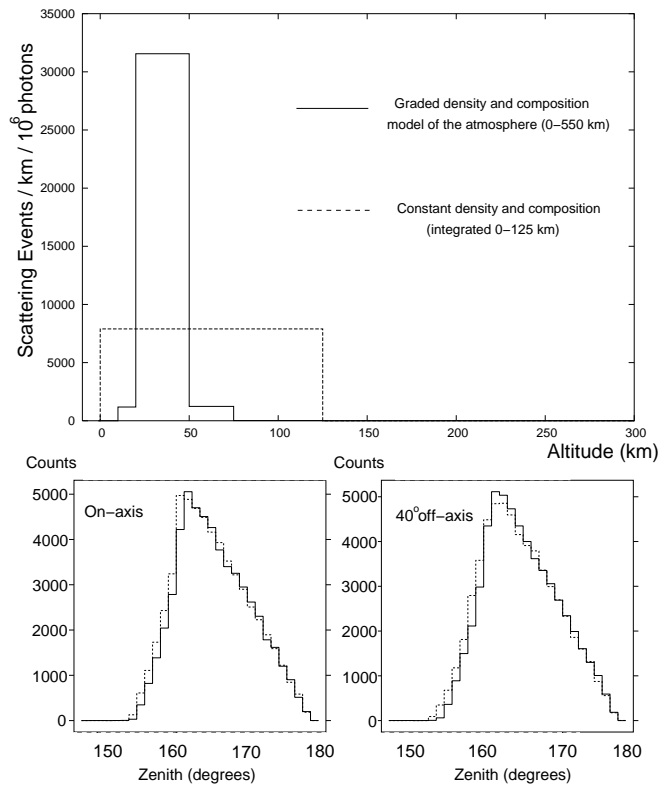}
\caption{A comparision between using a 12 layered graded atmospheric model and a constant density and composition integrated model. The incident photon energy is 200\,keV. The upper plot shows the back-scatter rate with altitude and the lower two histograms show the albedo zenith distribution.}
\label{fig:graddist}
\end{figure}

The model of the Earth's Atmosphere consists of a ball of constant density ($8.33 \times 10^{3}$\,g.cm$^{-3}$) and composition (38\% O$_{2}$ : 62\% N$_{2}$,). This is an approximation that will introduce a systematic error but aid in the simplicity and speed of BAPS. The constant density was acheived by averaging over the first 125 km of the atmosphere. To evaluate the effectiveness of this approximation, a single column element of this atmospheric model was compared to a graded (in density and composition) 12 layered model of the atmosphere. The rate of back-scattered events (scatter angle $>$90$^{\circ}$) and resultant direction of the photon emerging from the atmosphere were recorded for both models, as seen in Figure \ref{fig:graddist}. For the layered model, the majority of the scattering occurs in the 20--50 km region, and the 0--125km integration reproduces 99.7\% of these back-scattered events. The back-scattered zenith angle distribution for an on-axis beam of photons and one at 40$^{\circ}$ off-axis was also compared for the two models. There is a slight excess around the scatter angle of 160$^{\circ}$ and below for the approximated model. However, it is clear that this discrepancy between the two models will not effect the BAPS fitting when considering the uncollimated nature of the BATSE LADs. Furthermore, the 0.3\% change in the albedo count-rate will be accounted for in the fitting routine. The diametric excess that polarisation will induce will not be effected.

This depth-dependent scattering rate also discards the need for including the Earth itself into the model. A photon interacting such that it scatters in the atmosphere towards the direction of the \textit{CGRO} but must travel through the Earth to reach the \textit{CGRO} volume will certainly be absorbed by the atmosphere in the model, and so discarded.

\subsection{Estimating the Degree of Polarisation}
\label{sec:estimate}
\begin{figure}[ht]
\centering
\includegraphics[width=9cm]{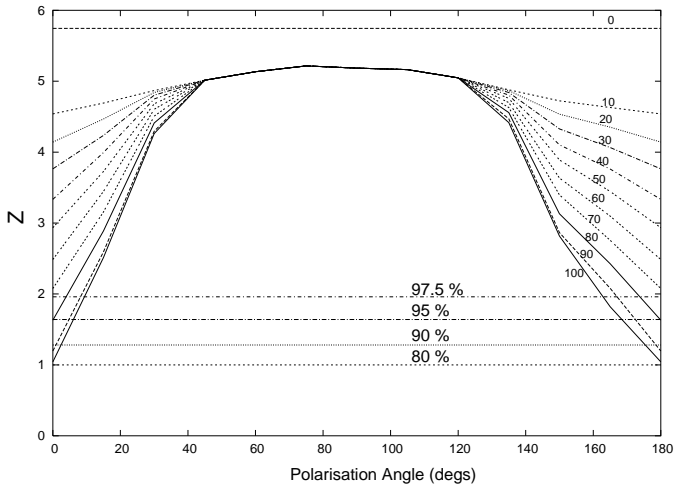}
\caption{The CONT data for GRB\,930131 is compared to the simulations with $\Pi$ restricted to a range of maximum values between 0\% and 100\%. The confidence levels in the Z statistic are shown.}
\label{fig:2151per}
\end{figure}

\begin{figure}[ht]
\centering
\includegraphics[width=9cm]{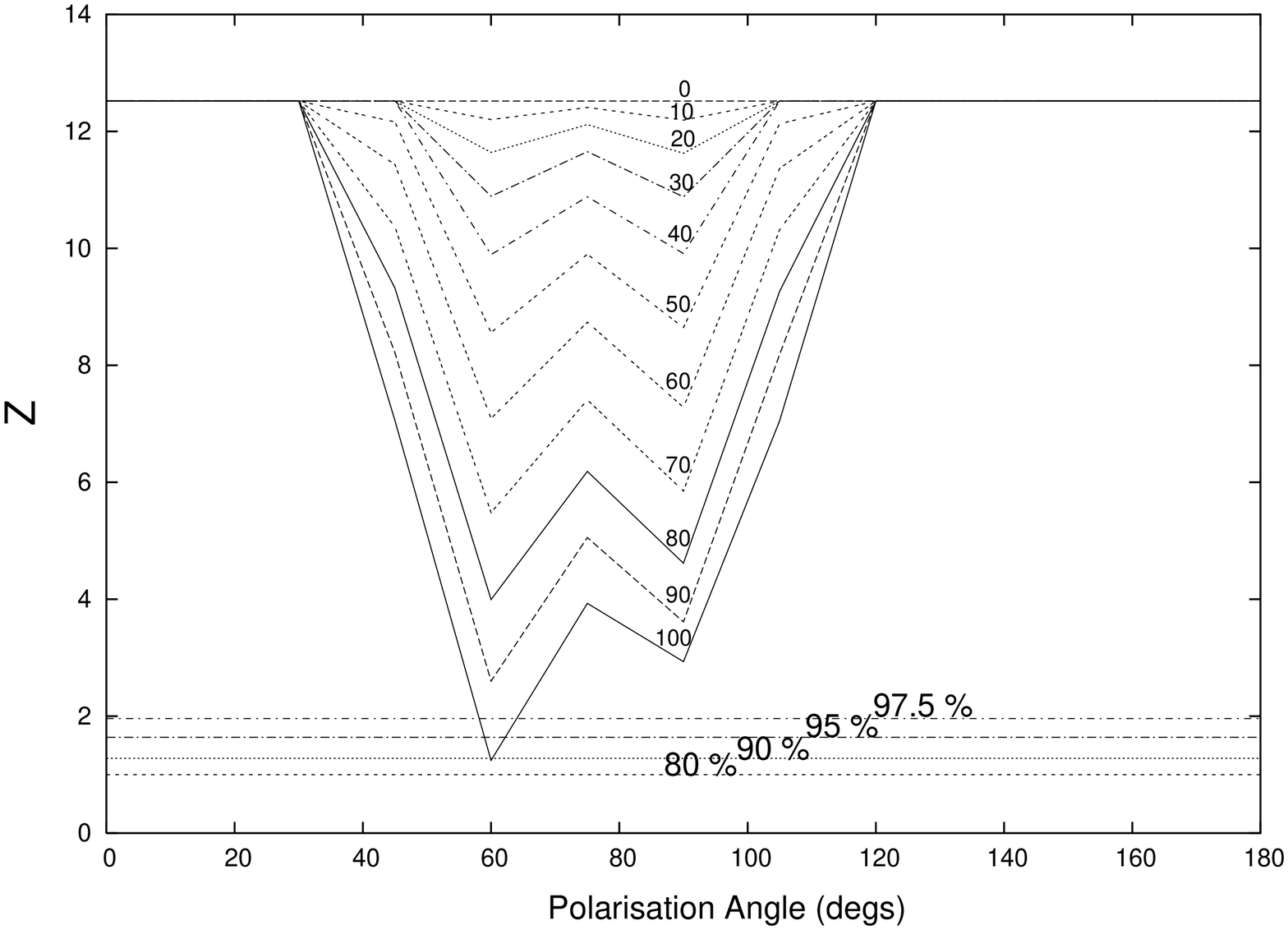}
\caption{The CONT data for GRB\,960924 is compared to the simulations with $\Pi$ restricted to a range of maximum values between 0\% and 100\%. The confidence levels in the Z statistic are shown.}
\label{fig:5614per}
\end{figure}

The physical implications of polarised prompt flux from GRBs hinges on $\Pi$, the percentage of the flux that is polarised. As BAPs normalises the levels of albedo flux and involves comparisons of count-rates, the degree to which these GRBs are polarised cannot be measured directly. However, an estimate of $\Pi$ can be achieved through restricting $\Pi$ to a maximum value and inspecting the change in the Z fitting statistic. The data is refitted with restrictions applied to $\Pi$, as shown in Figures \ref{fig:2151per} and \ref{fig:5614per}.

Though the significance of the fit for GRB\,930131 decreases as the maximum allowed value for $\Pi$ decreases, the fit is still significant down to a $\Pi$ values of around 70\%--80\%. However, the fit for GRB\,960924 suffers more severely with any change to $\Pi$ and the significance of the result is lost at a level of 90\% polarisation. This value for $\Pi$ is unphysical and so the systematics discussed in section \ref{sec:atmos} must be considered. As it is quite possible that the polarisation-induced modulation is underestimated by up to a factor of 2, the value of $\Pi$ may well be overestimated by a similar figure. Therefore the degree of polarisation in GRB\,930131 is consistent with $\Pi$ ranging between 35\% and 100\%; and GRB\,960624 is consistent with $\Pi$ ranging between 50\% and 100\%.

\section{Conclusions}
BAPS has been presented here in the context of GRB\,930131 and
GRB\,960924, both of which show a high degree of polarisation. The
caveats to this result have been discussed and no systematic
explanation can be found for these positive detections of
polarisation. The GRB selection procedure is initially highly
restrictive, however there are several bursts that are also suitable
to be analysed by BAPS in future studies. The initial stage of the
BAPS analysis (modelling
of the albedo flux) was run for a variety of angles taken from figure
\ref{fig:64} and it was seen that the difference
between the polarised and unpolarised cases was statistically significant up to $\eta < 120^{\circ}$, suggesting that BAPS is primarily
restricted in its sensitivity by the number of LADs that are
illuminated by the GRB albedo flux and not the off-axis angle of the burst. This increases the number
of GRBs that can be analysed with BAPS, however since the senstivity
is governed by the number
of illuminated LADs, it removes the possibilty of analysing the
systematics of a suitably off-axis GRB that is unable to exhibit any
polarisation effects.

There are areas in BAPS that would benefit from further augmentation: modification to the current GEANT4 release, an analytical model of the scattering off the Earth's atmosphere and a full analysis of the systematic errors to enable a quantitative estimate to the degree of polarisation and enable the inclusion of more LADs into the fitting proceedure.

Though the degree of polarisation is an estimate from these results, there is distinct evidence that polarisation exists in prompt GRB flux and to a degree that is large enough to be detected by viewing the Earth's atmosphere with the BATSE detectors. The implications of this on the GRB mechanism are that a large scale magnetic field is present in the emitting region, conclusions beyond this will remain uncertain until future measurements can be made. Viewing the prompt flux is crucial to future polarimetry measurements, something that is only currently possible through a large scale survey of the BATSE catalogue with BAPS or through an on-axis \textit{INTEGRAL} burst such as the recent strong burst GRB\,041219.

\begin{acknowledgements}
The author would like to acknowledge Colleen Wilson-Hodge and Valerie
Connaughton (MSFC) for their contributions from the BATSE side and
Christian Kaiser (Southampton) for general discussions.
\end{acknowledgements}

\bibliographystyle{aa}
\bibliography{2693bib}

\end{document}